\documentclass[a4paper,english]{lipics-v2019}
\usepackage{nth}
\usepackage{microtype}
\usepackage{breqn}
\usepackage{ifthen}
\usepackage{placeins}
\usepackage{complexity}
\usepackage{nicefrac}
\usepackage{todonotes}
\usepackage{siunitx}
\usepackage{amssymb}
\hideLIPIcs
\nolinenumbers

\newcommand{\old}[1]{{}}

\title{Minimum Partition into Plane Subgraphs:\\ The CG:SHOP Challenge 2022}
\titlerunning{Minimum Partition into Plane Subgraphs: CG Challenge 2022}

\author{Sándor P.~Fekete}{Department of Computer Science, TU Braunschweig, Germany}{s.fekete@tu-bs.de}{https://orcid.org/0000-0002-9062-4241}{}
\author{Phillip Keldenich}{Department of Computer Science, TU Braunschweig, Germany}{p.keldenich@tu-bs.de}{https://orcid.org/0000-0002-6677-5090}{}
\author{Dominik Krupke}{Department of Computer Science, TU Braunschweig, Germany}{d.krupke@tu-bs.de}{https://orcid.org/0000-0003-1573-3496}{}
\author{Stefan Schirra}{Department for Simulation and Graphics, OvGU Magdeburg, Germany}{stschirr@isg.cs.uni-magdeburg.de}{???TODO???}{}
\authorrunning{S.~P.~Fekete, P.~Keldenich, D.~Krupke, S.~Schirra}
\Copyright{S.~P.~Fekete, P.~Keldenich, D.~Krupke, S.~Schirra}



\supplement{}

\funding{Work at TU Braunschweig has been partially supported by the German Research Foundation (DFG), 
project ``Computational Geometry: Solving Hard Optimization Problems'' (CG:SHOP), grant FE407/21-1.} 

\acknowledgements{We thank the members of the CG Challenge Advisory Board for their valuable input: 
William J.~Cook, Andreas Fabri, Michael Kerber, Philipp Kindermann, Joe Mitchell, Kevin Verbeek.}

\EventAcronym{CG:SHOP 2022}

\begin{document}
\maketitle
\begin{abstract}
We give an overview of the 2022 Computational Geometry Challenge
targeting the problem {\sc Minimum Partition into Plane Subsets},
which consists of partitioning a given set of
line segments into a minimum number of non-crossing subsets.
\end{abstract}

\section{Introduction}
The ``CG:SHOP Challenge'' (Computational Geometry: Solving Hard
Optimization Problems) originated as a workshop at the 2019
Computational Geometry Week (CG Week) in Portland, Oregon in June,
2019.  The goal was to conduct a computational challenge competition
that focused attention on a specific hard geometric optimization
problem, encouraging researchers to devise and implement solution
methods that could be compared scientifically based on how well they
performed on a database of carefully selected and varied instances.
While much of computational
geometry research is theoretical, often seeking provable approximation
algorithms for \NP-hard optimization problems,
the goal of the CG Challenge was to set the metric of success based on
computational results on a specific set of benchmark geometric
instances. The 2019 CG Challenge focused on the problem of computing
simple polygons of minimum and maximum area for given sets of vertices in the
plane. This Challenge generated a strong response from many research
groups, from both the computational geometry and the combinatorial
optimization communities, and resulted in a lively exchange of
solution ideas.

For CG Weeks 2020 and 2021, the Challenge problems were {\sc Minimum Convex Partition}
and {\sc Coordinated Motion Planning}, respectively. The CG:SHOP Challenge became an event within
the CG Week program, with top performing solutions reported in the
Symposium on Computational Geometry proceedings. The schedule for the
Challenge was advanced earlier, to give an opportunity for more
participation, particularly among students, e.g., as part of course
projects. 

The fourth edition of the Challenge in 2022 continued
this format, leading to contributions in the SoCG proceedings.
A record number of 40 teams registered, with 32 submitting at least one valid solution.

\section{The Challenge: Partition into Plane Subgraphs}

A suitable contest problem has a number of desirable properties.

\begin{itemize}
\item The problem is of geometric nature.
\item The problem is of general scientific interest and has received previous attention.
\item Optimization problems tend to be more suitable than feasibility problems; in principle, 
  feasibility problems are also possible, but they need to be suitable for sufficiently
  fine-grained scoring to produce an interesting contest.
\item Computing optimal solutions is difficult for instances of reasonable size.
\item This difficulty is of a fundamental algorithmic nature, and not only due to
 issues of encoding or access to sophisticated software or hardware.
\item Verifying feasibility of provided solutions is relatively easy.
\end{itemize}

In this fourth year, a call for suitable problems was communicated in May 2021.
In response, a total of seven interesting problems were proposed for the 2022 Challenge.
These were evaluated with respect to difficulty, distinctiveness from previous years,
and existing literature and related work. In the end, the Advisory Board
selected the chosen problem. Special thanks go to Johannes Obenaus (FU Berlin)
who suggested this problem, based on his own related work~\cite{obenaus2021edge,aichholzer2020plane}.

\subsection{The Problem}
\begin{figure}
	\centering
	\includegraphics{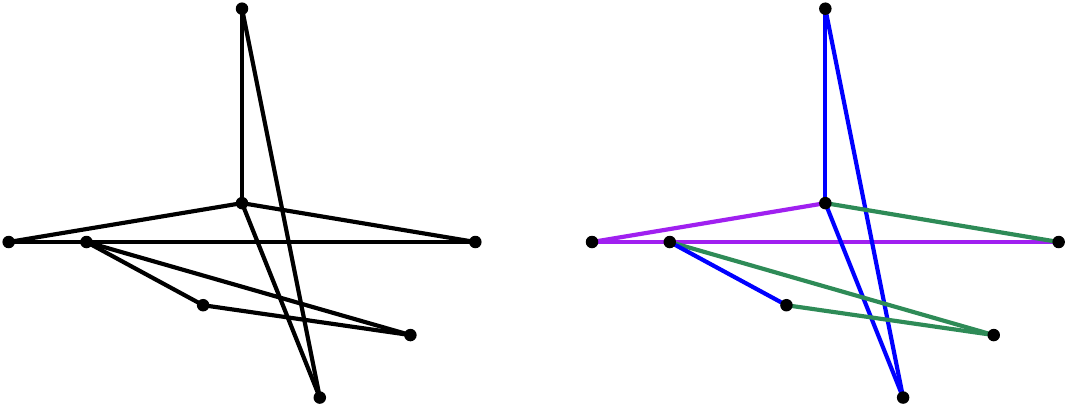}
	\caption{A possible instance, given by a set of line segments in the plane (left)
			 and a feasible partition into three plane subgraphs
			 represented as an edge coloring (right).
			 Note that segments sharing a common endpoint are not considered 
			 intersecting, but a segment with a point in the interior of another segment
			 is considered to intersect that segment.}
	\label{fig:example}
\end{figure}

The specific problem that formed the basis of the 2022 CG Challenge was the following;
see Figure~\ref{fig:example} for a simple example.

\medskip
\noindent \textbf{Problem:} {\sc Partitioning into Plane Subgraphs} 

\noindent \textbf{Given:} A set of $n$ distinct segments $E = \{s_1,\ldots,s_n\}$ in the plane.
Arbitrarily many segments may share a common endpoint and the points need not be in general position.
This input can also be considered to be the straight-line embedding of a graph $G = (V,E)$
with a vertex for each distinct segment endpoint and an edge corresponding to each segment.

\noindent \textbf{Goal:} The task is to compute a partition of $E$ into a minimum number $k$ of
disjoint subsets $S_1,\ldots,S_k \subseteq E$, such that $\bigcup_{i=1}^{k} S_i = E$
and no two segments $s_j,s_{\ell} \in E_i$ in the same subset $E_i$ intersect.

For the purpose of this contest, two segments are considered intersecting
iff they have a common point that is not an endpoint of both segments.
In other words, two segments intersect if they share a point that is an endpoint of neither segment
or if the endpoint of one segment lies in the interior of the other.
By this definition, segments that merely share an endpoint do not intersect.

The problem can also be cast as graph coloring on the
\emph{conflict graph} (also known as \emph{intersection graph}) of $E$.
This graph $G' = (V',E')$ contains a vertex for each segment, i.e., $V' = E$,
and two vertices are adjacent in $G'$ if the corresponding segments intersect.
In graph coloring, the task is to assign a color value from $\{1,\ldots,k\}$ to
each of the vertices, such that adjacent vertices receive different colors,
while minimizing the number of colors $k$.

\subsection{Related Work}
There is not much research regarding the specific, geometric version of the contest problem.
However, the fact that it can be equivalently stated as graph coloring of the conflict graph
means that any research on solving graph coloring can be applied directly.
Due to its fundamental nature and its prominence in theoretical computer science and mathematics,
there is more research on graph coloring than we can reasonably cite here.
We therefore focus on practical approaches that do not require graphs to have certain attributes
and that could be applied to our problem; an overview over many heuristic approaches
can also be found in~\cite{jensen2011graph} or in the book by Lewis~\cite{lewisbook},
who also provides implementations of several heuristic and metaheuristic approaches.
Additionally, much of the existing practical research on graph coloring focuses
on rather sparse graphs~\cite{10.1007/978-3-030-19212-9_25, lin2017reduction}.
In contrast, the conflict graphs in our contest are typically fairly dense with somewhere between
\SI{1.8}{\percent} and \SI{62.5}{\percent} of all possible edges;
some of them have more than 1.5 billion edges with under 75,000 nodes.

\subsubsection{Simple Heuristics}
In order to come up with a decent initial coloring, greedy heuristics can be used.
A very well-known greedy heuristic due to Welsh and Powell~\cite{welsh1967upper}
assigns the lowest possible color to the uncolored vertex with highest degree,
until all vertices are colored.

This result has been improved in various ways.
One of the most prominent greedy coloring strategies is called DSATUR (degree of saturation), and is due to Brélaz~\cite{dsatur}.
A suitable implementation, in particular for dense graphs, runs in $\mathcal{O}(n^2)$ time.
It is also possible to achieve an $\mathcal{O}((n+m)\log n)$ running time, which may be preferable
if the number of edges $m$ is $o(\nicefrac{n^2}{\log n})$.
It assigns colors to vertices one-by-one, like the classic greedy algorithm,
but instead of choosing the vertex with maximum degree, it chooses the vertex with most colors,
breaking ties by maximum degree.
This heuristic has been studied w.r.t. various aspects, including theoretical guarantees for
certain graph classes and sizes~\cite{janczewski2001smallest}.

Other greedy-type heuristics include RLF due to Leighton~\cite{rlf}, which runs in time $\mathcal{O}(nm)$.
It constructs one maximal color class at a time, while trying to minimize the number of edges
that remain in the uncolored graph, i.e., edges that may still cause conflicts in each step.
Variations of this heuristic have been studied and shown to be successful for certain graph classes~\cite{matula1972graph}. 

\subsubsection{Exact Methods}
DSATUR has also been used as a subroutine in a branch-and-bound-based
exact graph coloring algorithm, sometimes called Backtracking DSATUR~\cite{sager1991pruning},
which became the de-facto standard algorithm for exact graph coloring
for a while after its presentation in 1991; it was later improved 
by further research, in part driven by the 1993 DIMACS implementation challenge on
Cliques, Coloring and Satisfiability~\cite{sewell1996improved}.
More recently, San Segundo~\cite{san2012new} and Furini et al.~\cite{furini2017improved}
presented improved branch-and-bound-based exact coloring methods
that also make extensive use of DSATUR.

Graph coloring has also been tackled using branch-and-bound
algorithms based on integer programming~\cite{mendez2006branch}.
The obvious formulation (with a zero-one-variable for each vertex and each color)
is known to have a weak relaxation
that suffers from many symmetric solutions due to
the fact that colors can be used interchangeably, and
has trivial solutions with value 2 for any graph~\cite{malaguti2010survey}.

Another formulation proposed by Mehrotra and Trick~\cite{mehrotra1996column}
turns the coloring problem into a set cover problem by introducing
binary variables for independent sets in a column-generation approach,
aiming to cover all vertices with as few independent sets as possible.
This approach was adapted into a branch-cut-and-price approach by
Furini and Malaguti~\cite{furini2012exact}; it can also be based on constraint programming
instead of integer programming~\cite{gualandi2012exact}.
These approaches are typically enough to solve instances with several hundred vertices;
the size of our contest instances was significantly higher, consisting of at least
several thousand vertices in the conflict graph.

Other approaches to exact graph coloring involve listing small independent sets~\cite{eppstein2002small}
or make use of tree-, path- or linear-decompositions of a graph~\cite{lucet2006exact}.

\subsubsection{Metaheuristics \& Improvement Search}
Due to the limited size of instances that can still be solved to provable optimality,
most practical methods focus on providing good solution without aiming for provable optimality.
These methods typically start with an initial coloring produced by the application of one or more simple heuristics,
such as DSATUR or RLF,
and then try to improve it.

One way to improve a given initial coloring is by using a folklore procedure sometimes called \emph{iterated greedy coloring}.
Given a valid coloring, permute the color values;
this can be done according to different strategies,
e.g., by reversing the color values or by randomly permuting them.
Afterwards, iterate through the vertices in order of non-decreasing color,
and assign to each vertex the lowest color that is not currently present in its neighborhood.
Because each vertex will receive at most its color in the permuted list of colors,
the resulting coloring will use at most the same number of colors as the original coloring,
but may also use fewer colors.
Repeated application of this scheme tends to reduce the number of colors for colorings produced by simple heuristics.

While being fast and simple, this approach, which maintains the validity of the coloring in each step,
is typically weak when compared to other approaches to improving a coloring.
These typically try to remove a color class, e.g.,
by recoloring all vertices with random colors.
This of course may make the coloring invalid;
typical improvement methods then try to restore the validity of the coloring
by minimizing the number of conflicts --- i.e., the number of edges $vw$ with $c(v) = c(w)$ ---
using some type of local search, ultimately trying to reduce it to 0.

One classical approach in this vein is the TABUCOL algorithm introduced by Hertz and De Werra~\cite{hertz1987using}, which tries to minimize the number of conflicts using tabu search.
The basic operation is to take a conflicting vertex and recolor it, such that the number of conflicts is minimized, subject to a tabu list of color assignments.

Another local search approach is to allow partial colorings instead of conflicts;
removing a color class is done by making vertices from some color class uncolored.
This approach was introduced by Blöchliger and Zufferey~\cite{blochliger2008graph}
for their PARTIALCOL algorithm, which also makes use of tabu search to minimize
the number of uncolored vertices.

Another approach to minimizing the number of conflicts using metaheuristics is
the use of population-based algorithms.
They are often combined with tabu search or other local search mechanisms,
which can be applied in a population-based algorithm in several ways,
into so-called \emph{hybrid evolutionary algorithms},
yielding some of the best recently developed coloring metaheuristics~\cite{moalic2018variations}.
For instance, tabu search may be used as a mutation operator as
in MACOL~\cite{lu2010memetic}; it may also simply be applied on each new population member
to restore a decent level of conflict-freeness after crossover.

There are also multiple options for mate selection and crossover.
One of the best-known crossover routines is called \emph{Greedy Partition Crossover} (GPX)
and due to Galinier and Hao~\cite{galinier1999hybrid}.
In it, a possibly conflicting $k$-coloring is constructed from
two parents by carrying over entire color classes from a parent to the child,
relabeling colors in the process.
The next color class to be copied is chosen greedily,
maximizing the number of currently uncolored vertices in the class.
The first color assignment to a vertex done in this manner determines
the color of the vertex in the child.
Vertices that remain uncolored --- because only $k$ color classes can be copied --- are assigned
a color in some way, e.g., randomly or using a heuristic such as DSATUR.
Other crossover operators choose the color class to be copied
randomly~\cite{falkenauer1996hybrid} or form new color classes from the union of color
classes of the parents~\cite{dorne1998new}; in more recent work,
crossover has become even more complex~\cite{porumbel2010evolutionary}.
In recent work, Goudet et al.~\cite{goudet2021deep} combine deep learning with 
population-based metaheuristics to obtain a new coloring algorithm that they find
highly competitive compared to existing metaheuristics.

Another approach that is different from simple local search procedures and
hybrids between population-based and local-search-based metaheuristics
is \emph{quantum annealing}~\cite{titiloye2011quantum, titiloye2012parameter}.

In recent years, research has also focused on computing colorings for very large graphs in
a distributed fashion; in many cases, these approaches are applied to graphs that are too
large even for the standard greedy heuristics, which is not the case for our contest graphs.
These methods often begin by stripping a graph of its low-degree vertices,
and then trying to solve the coloring problem on the remaining kernel.
Hebrard and Katsirelos~\cite{10.1007/978-3-030-19212-9_25} present an algorithm
based on a combination of the idea outlined above with local search.
They present methods to find subgraphs that require a large number of colors,
which provides them with good lower bounds, allowing them to cut off more vertices;
on the other hand, a local search procedure is used to find good colorings.
In some cases, this can lead to provably optimal results for large graphs.

\subsection{Instances}
An important part of any challenge is the creation of suitable instances.
If the instances are easy to solve to optimality, the challenge becomes trivial;
On the other hand, if instances require a huge amount of computation for important common pre-processing steps
or for finding any decent solutions, the challenge may heavily favor teams that can afford better computation equipment.
The same is true if the set of instances becomes too large to manage with a single (or few) computers.

We implemented simple exact solvers based on integer programming, constraint programming,
SAT solvers and heuristics to obtain a rough estimate on the maximum instance size
that can be solved to optimality using these methods.
SAT solvers were able to solve some instances exceeding 2,000 segments in a day;
therefore, our contest did not contain any instances with fewer than 2,000 segments.

A priori, it is difficult to tell how hard finding a good solution to an instance is and which parameters
influence the difficulty of solving an instance (and in what way).
Therefore, it is important to create a set of instances that are diverse with respect to parameters that are
likely to influence their difficulty.

Furthermore, it can be argued that good instances should also avoid
containing too many parts that can be easily discarded.
They also should not offer ways to subdivide the problem into independent subproblems.
If the instances have many such features, they may become much easier than anticipated
by their size or may even become accidentally trivial.

In the context of this contest, this means that the conflict graphs should be connected
and should not easily be disconnected by removing a few vertices.
Furthermore, they should not have too many vertices of small degree:
any vertex with degree $\deg(v) < k$ for some lower bound $k$ on the required number of colors 
can simply be discarded; such vertices can later be colored in a greedy fashion
in reverse order of removal without increasing the number of colors.
Similarly, if two (non-adjacent) vertices $v, w$ have $N(v) \subseteq N(w)$,
$v$ can be removed and later be colored using the same color as $w$.
Such vertices increase an instance's size without contributing to its difficulty.

Besides the aforementioned issues, the presence of a large fraction of trivially removable features
could force contestants to implement an even larger number of well-known ideas in their code
instead of focusing on developing fresh ideas.
Of course it can also be argued that the removal of such features should not be done,
e.g., because practical instances can also have trivially discardable parts or
may admit a subdivision into independent subproblems,
and making use of such features is important in a solver that is used on real-world instances.
It is therefore a judgement call whether to extensively prune such features from
contest instances during their generation or not.
We did not extensively apply such pruning techniques to our instances, but
for several generators we took care to avoid trivial instances and situations
in which a subdivision into independent subproblems was obvious, such as disconnected conflict graphs.

\begin{figure}
	\centering
	\begin{subfigure}[b]{.32\textwidth}
		\centering
		\includegraphics[width=\textwidth]{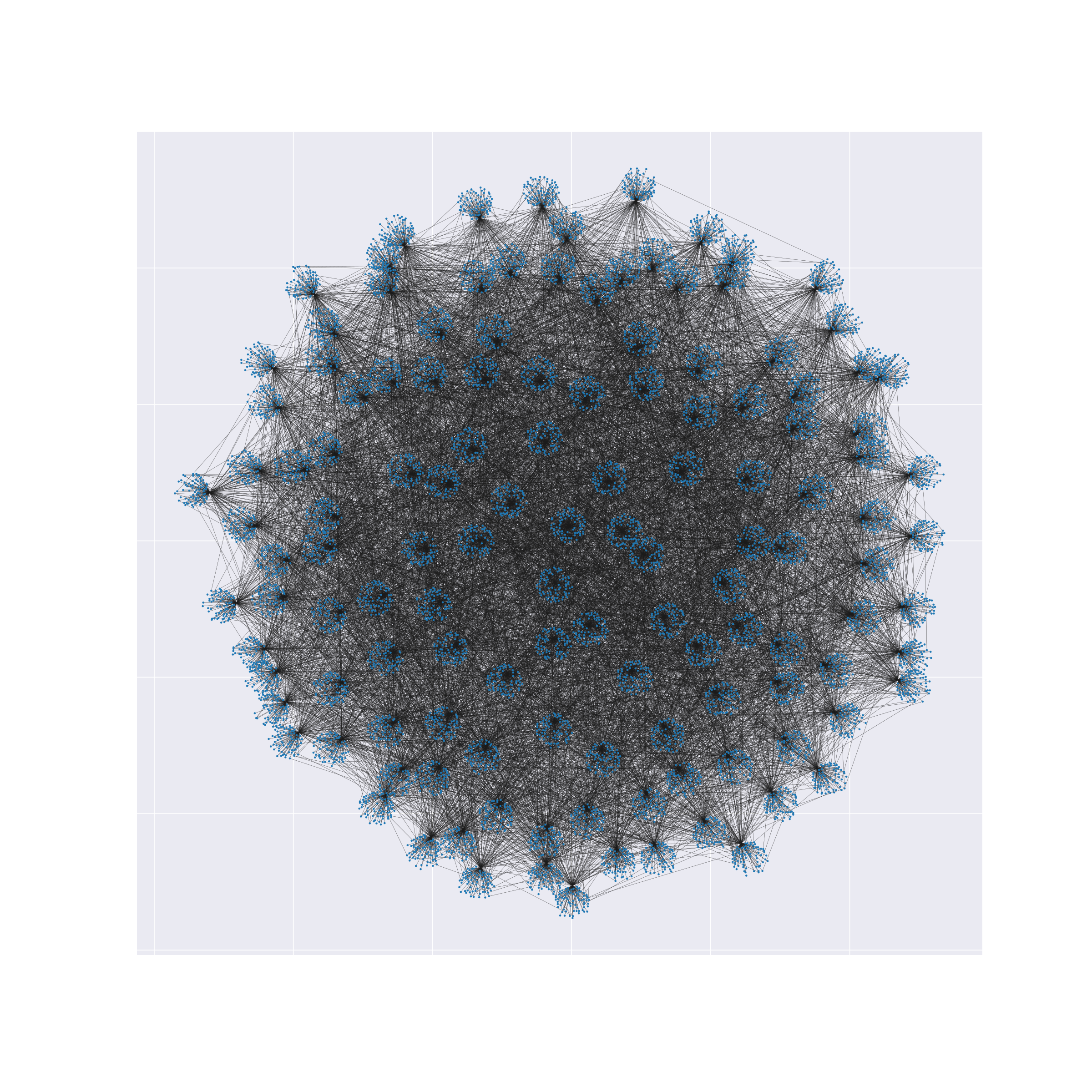}
		\caption{Instance \texttt{reecn6910}.}
	\end{subfigure}\hfill
	\begin{subfigure}[b]{.32\textwidth}
		\centering
		\includegraphics[width=\textwidth]{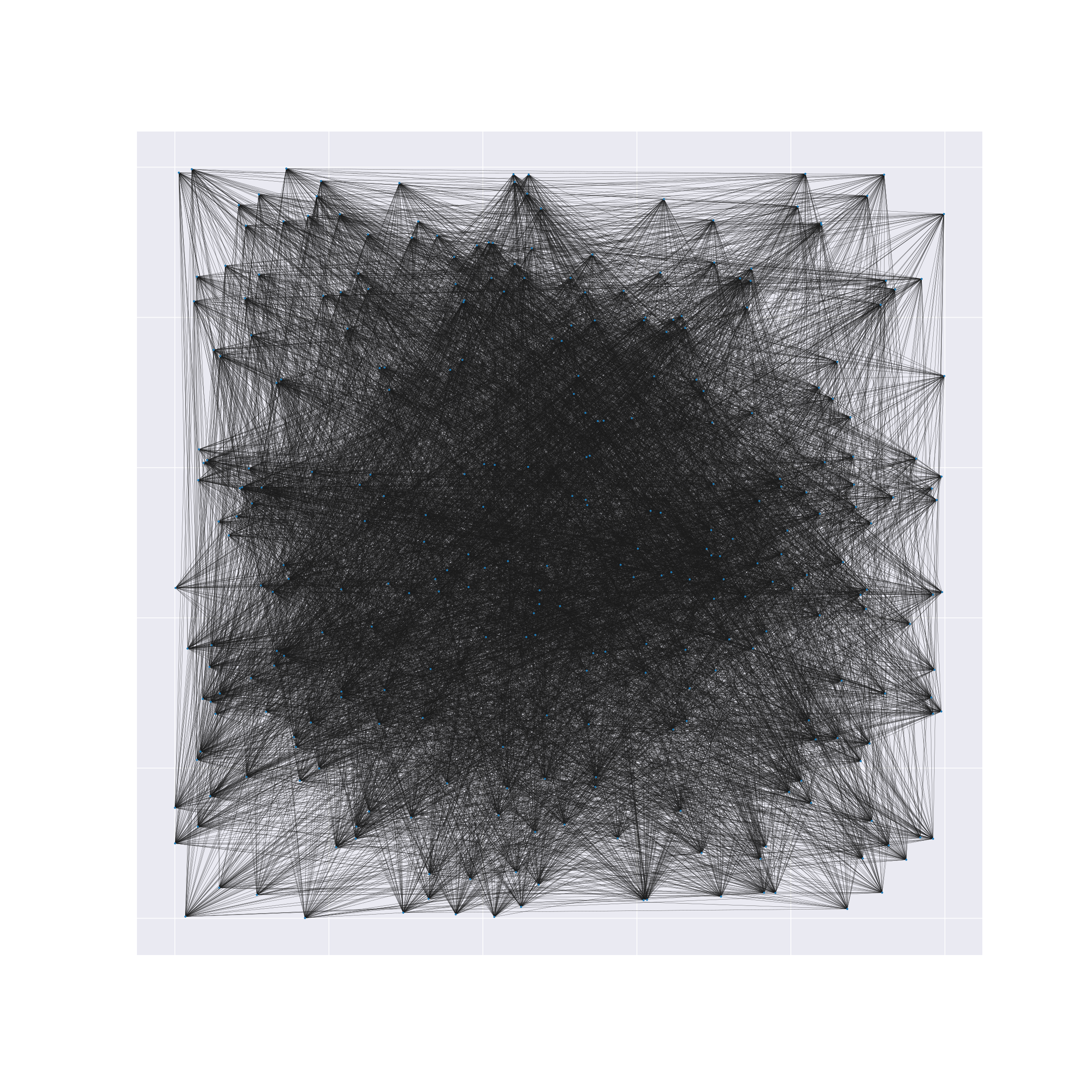}
		\caption{Instance \texttt{sqrp7730}.}
	\end{subfigure}\hfill
	\begin{subfigure}[b]{.32\textwidth}
		\centering
		\includegraphics[width=\textwidth]{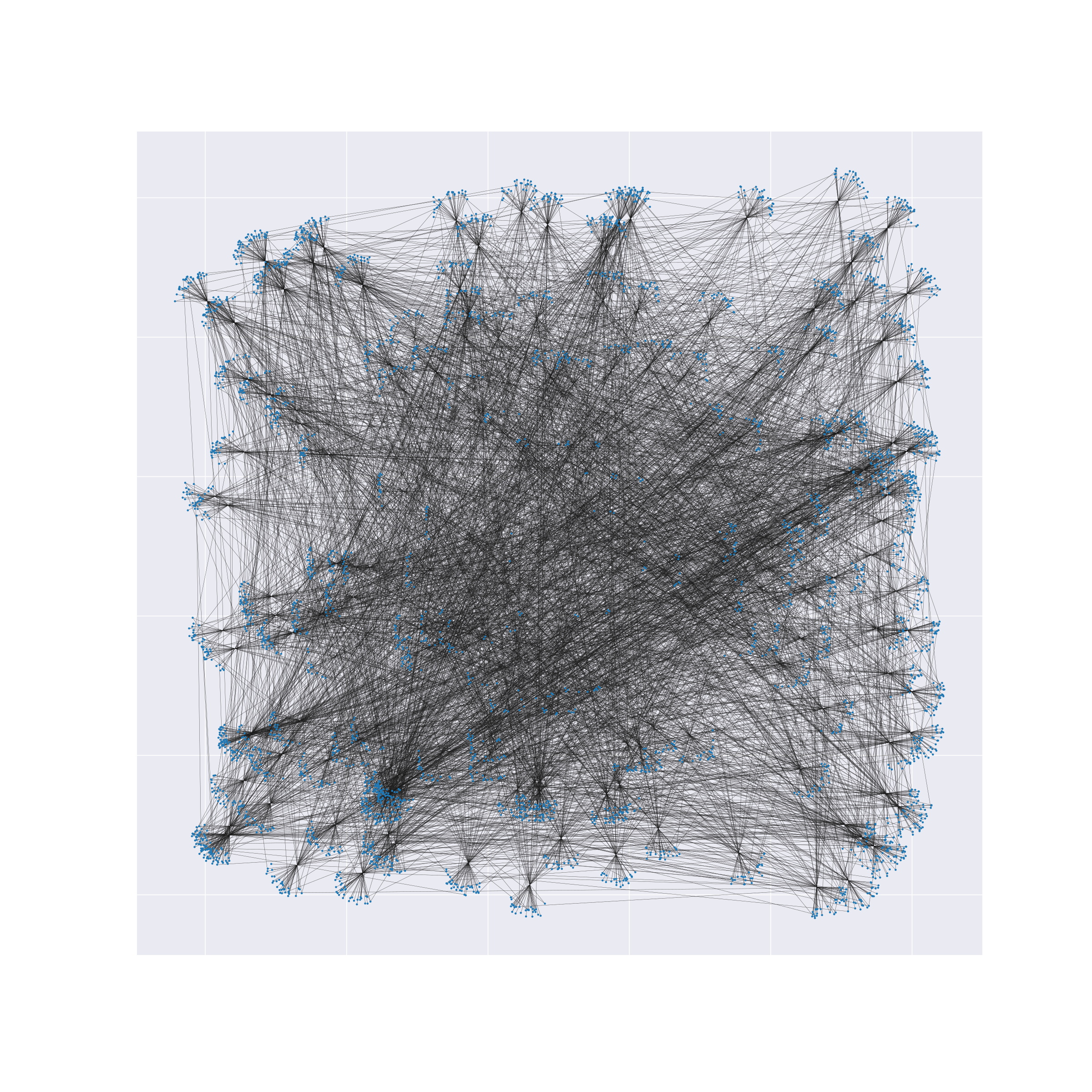}
		\caption{Instance \texttt{sqrpecn3020}.}
	\end{subfigure}
	\begin{subfigure}[b]{.32\textwidth}
		\centering
		\includegraphics[width=\textwidth]{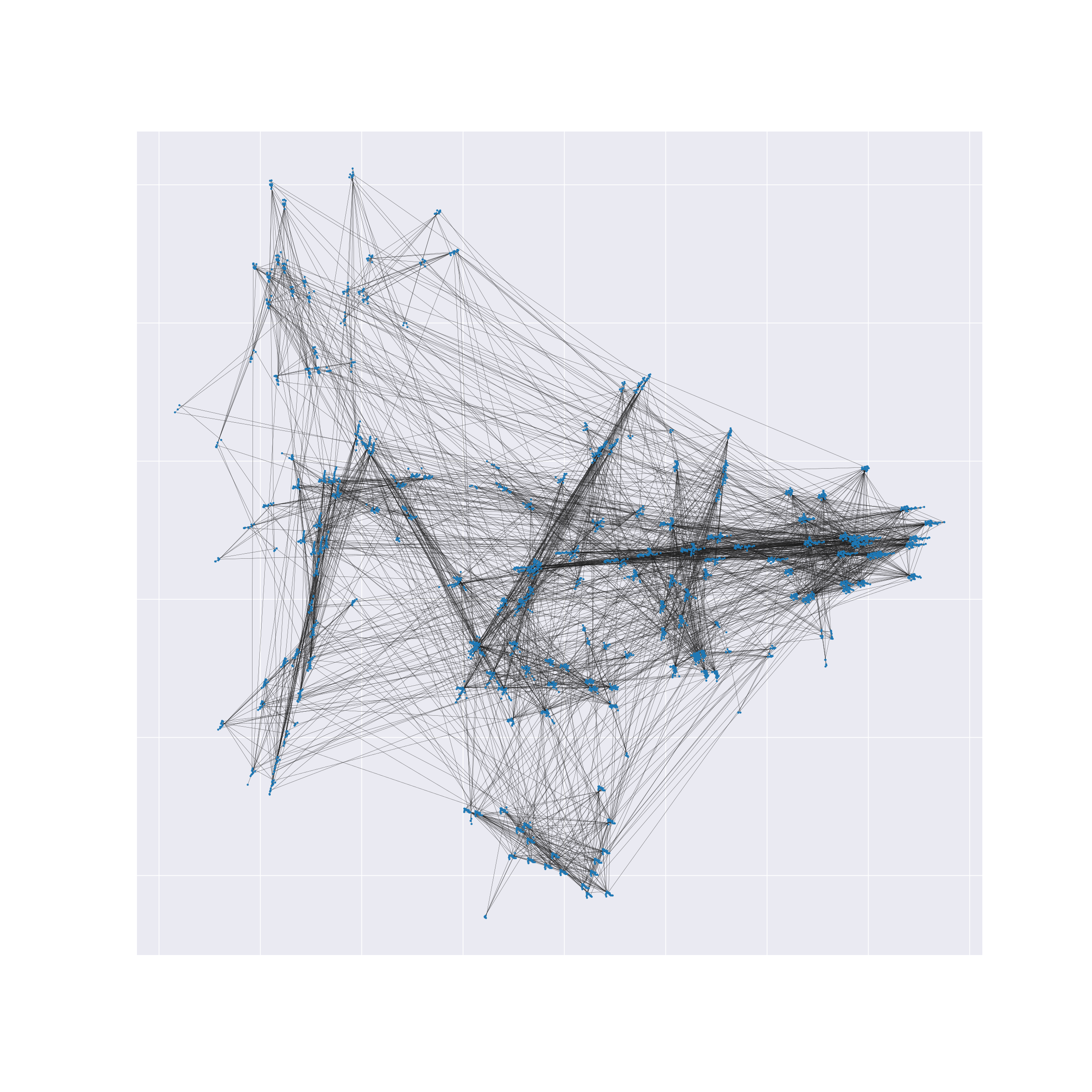}
		\caption{Instance \texttt{vispecn2518}.}
	\end{subfigure}\hfill
	\begin{subfigure}[b]{.32\textwidth}
		\centering
		\includegraphics[width=\textwidth]{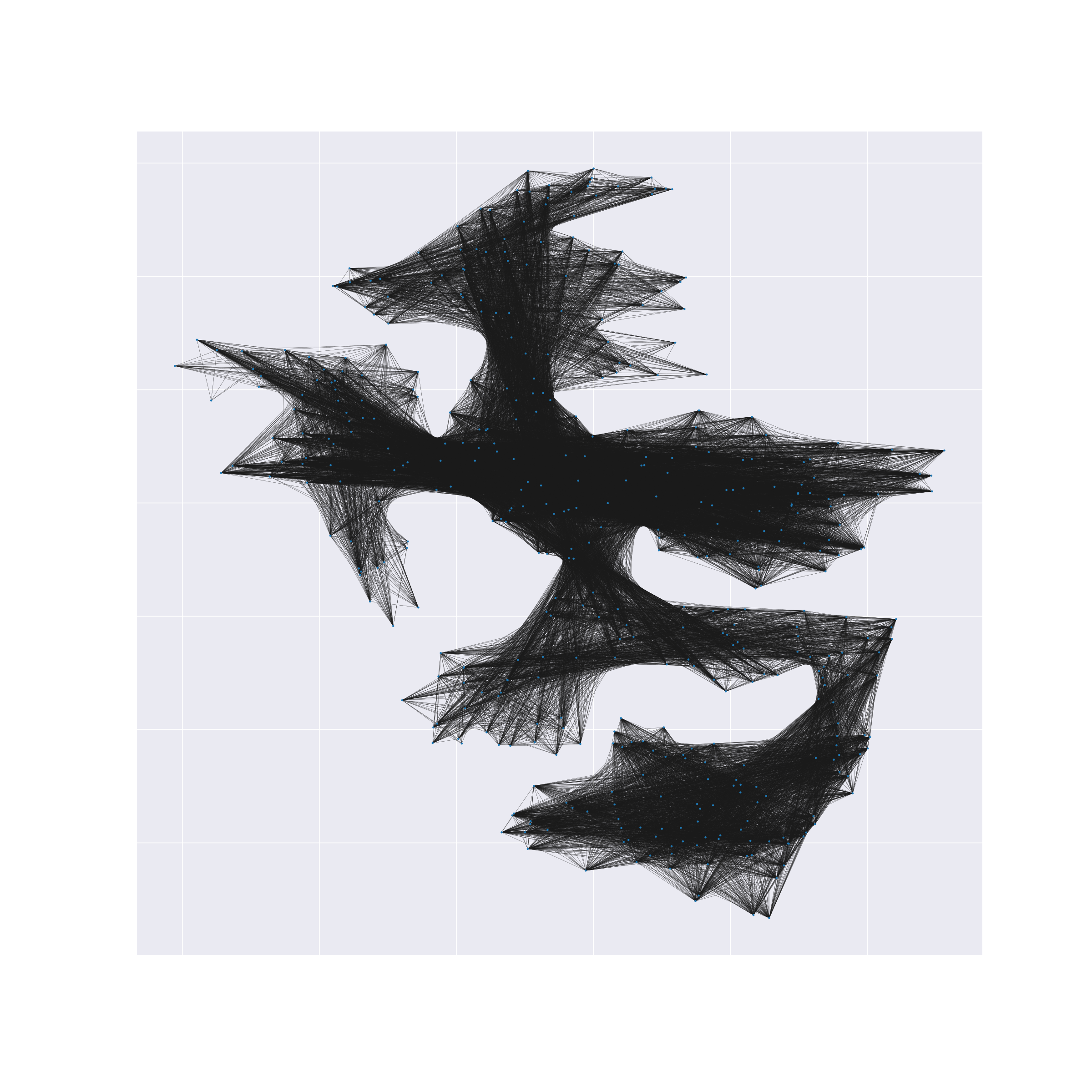}
		\caption{Instance \texttt{rvisp20601}.}
	\end{subfigure}\hfill
	\begin{subfigure}[b]{.32\textwidth}
		\centering
		\includegraphics[width=\textwidth]{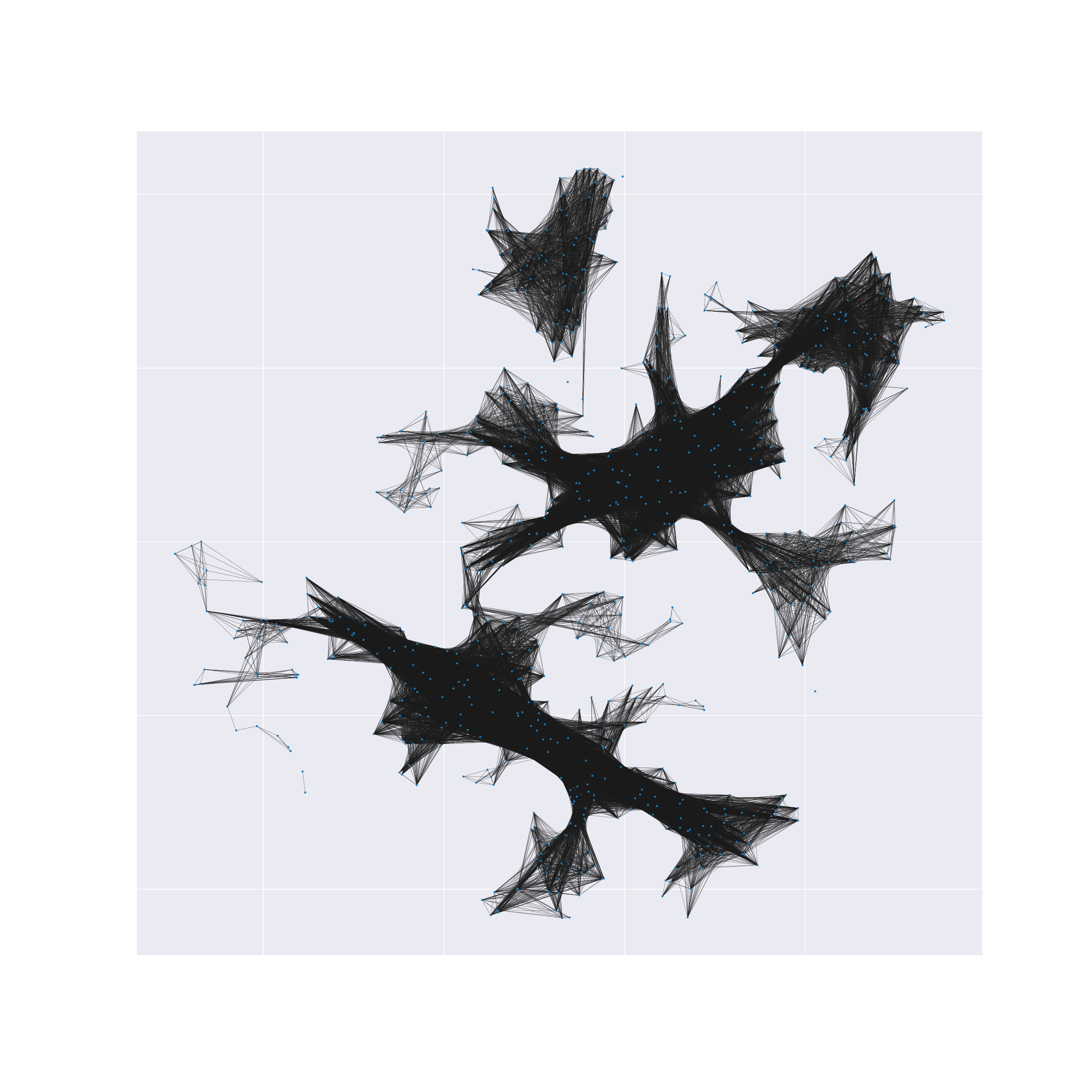}
		\caption{Instance \texttt{visp26405}.}
	\end{subfigure}
	\caption{A selection of actual contest instances made by different generators.}
	\label{fig:instance-examples}
\end{figure}

To create interesting and challenging instances, we developed several instance generators
that can be tuned by adjusting several parameters to create diverse instances.
We used the following generators to generate our instances.
\begin{description}
	\item[sqrp]
		We begin by generating a set of integer points $P$,
		choosing integer $x$- and $y$-coordinates uniformly at random
		within some rectangle $[0, a] \times [0, b]$.
		We then compute the list of all possible edges on $P$ and
		sort them by their length.
		We then exclude all edges that are outside some \emph{quantile range},
		i.e., edges that are shorter than $q$ percent or longer than $r$ percent
		of all edges.
		We then uniformly sample our instance's edges from the remaining edges,
		adding each edge uniformly at random with some probability $p$.
	\item[visp]
		We begin by obtaining a polygon $\mathcal{D}$ from the
		Salzburg Database of Polygonal Data~\cite{salzburgpoly} 
		with a number of vertices within a certain range.
		We then randomly generate a set of $m$ points $P$ in the interior of that polygon
		uniformly at random and connect each pair $p,q$ of points by an edge
		if $p$ is visible from $q$ in $\mathcal{D}$.
		The connected points form the segments of our resulting instance.
		Should the conflict graph of the instance be disconnected into several large components,
		or the instance be too large, or too small, we discard it and retry.
	\item[re]
		We begin by generating a random regular graph with some number $\ell$ of vertices 
		and some regular degree $m$, such that $m\ell$ is even.
		We then embed this graph in the plane,
		using the algorithm of Kamada and Kawai~\cite{kamadakawai}.
		The edges of this embedded graph are included in the instance.
		Finally, we add some additional edges uniformly at random between
		the points with a small probability $p$.
\end{description}
Each generator is optionally modified by a procedure that adds an aspect of edge coloring
to the instance. In edge coloring, we have to color edges that share a vertex with different colors.
Our procedure scales each individual segment by a factor $1 + \varepsilon$,
for some small $\varepsilon$, shifting its endpoints outwards from the segment's center.
In this way, segments that formerly shared an endpoint and could potentially receive the same color
are now truly intersecting and must receive different colors.
We denote the application of this procedure by a suffix \texttt{ecn} added to the
generator name.
Furthermore, each generator can be augmented by a procedure that adds a small number of edges
between randomly selected points as noise.
We denote the application of this procedure by a prefix \texttt{r} added to the generator name.
Our contest instances used the generators \texttt{sqrp}, \texttt{sqrpecn},
\texttt{visp}, \texttt{vispecn} and \texttt{reecn}.
In all our generators, we maintain integrality of coordinates
by scaling with a large constant and rounding the resulting coordinates to the nearest integer.
Figure~\ref{fig:instance-examples} shows examples of actual contest instances.

We measured several properties for each of the generated instances, including 
the number of edges,
the density, i.e., the number of edges compared to the ${|V| \choose 2}$ potential edges,
the number of edges and density of the conflict graph, as well as
the number of colors required by well-known coloring heuristics
such as DSATUR~\cite{dsatur} and RLF~\cite{rlf}.

\begin{figure}
  \centering
  \includegraphics[width=.98\textwidth]{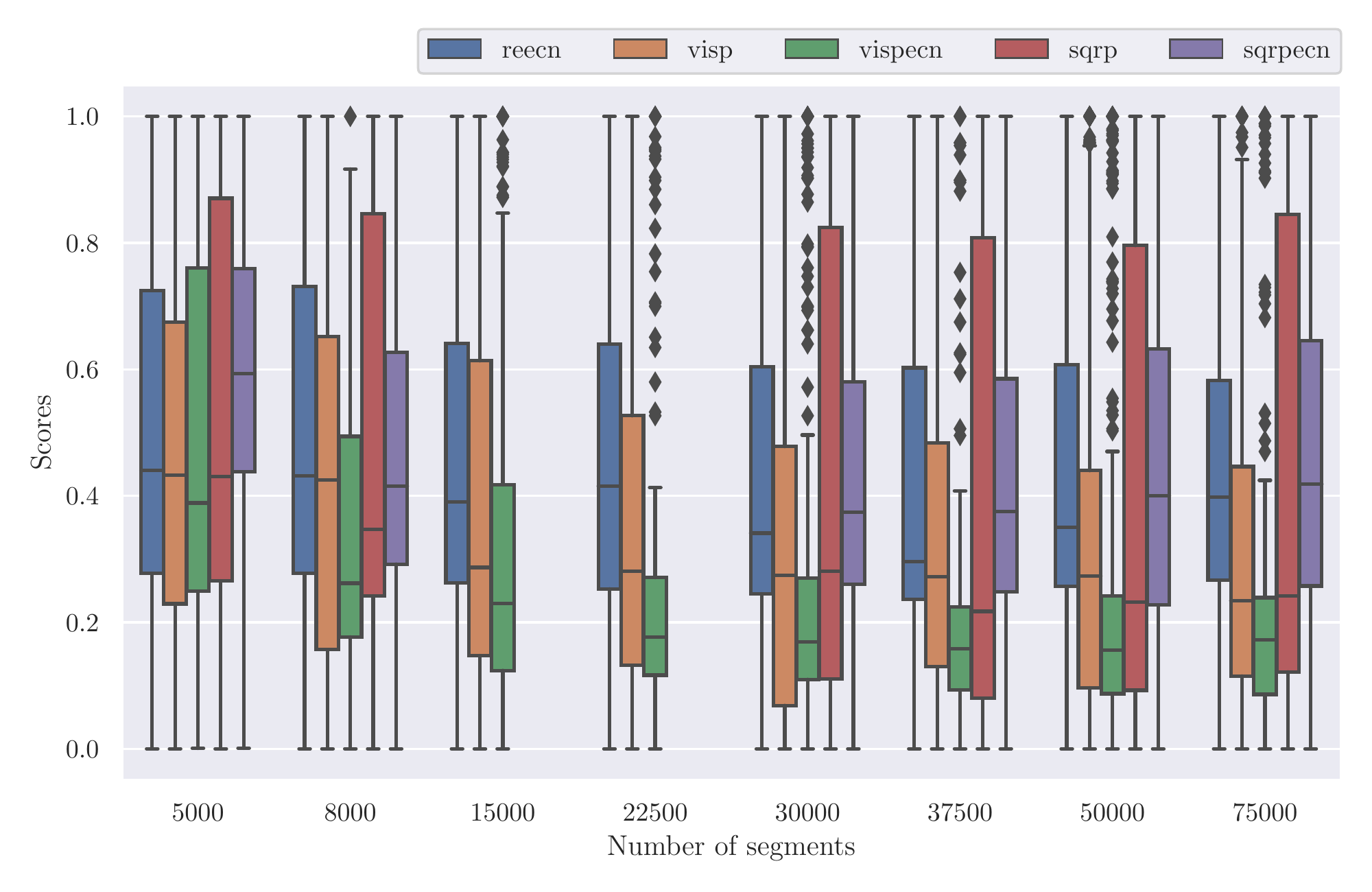}
  \caption{The distribution of the scores achieved by the teams, by instance type and size.
	       This plot provides some clues for the practical difficulty of instances, indicated
		   by low average scores.
		   Because the scores are based on the best submitted solution, this
		   implies that some teams with special approaches where able to obtain
		   significantly better results than ``basic'' approaches.
		   In particular, \texttt{vispecn} instances seem to be harder
		   for most participants than many other instance types.}
	\label{fig:hardness}
\end{figure}

Based on these properties, we then defined and tuned a distance function between the instances and used a
greedy dispersion algorithm to select a total of 225 diverse instances from a much larger set of candidate instances;
see Figure~\ref{fig:hardness} for an analysis of the difficulty of the various instances.

\subsection{Evaluation}
The contest was run on a total of 225 instances.
For the coloring problem, it is often considerably harder to find 
a solution with the optimal number of colors $k$ than it is to find a solution
that comes close to the optimum, say $k+1$ or $k+2$ colors.

In order to reflect this in the scoring of solutions,
instead of a score that linearly depends on the number of colors, 
we introduced an exponential scoring function.
For an instance $I$, let $B(I)$ be the number of colors
required in the best solutions submitted for that instance.
Furthermore, let $T(I)$ be the number of colors in the best solution of some team $T$ for $I$.
The score $S_T(I)$ of team $T$ for instance $I$ is
\[ S_T(I) := 0.95^{100 \cdot \frac{T(I)-B(I)}{B(I)}}. \]
In other words, the team loses \SI{5}{\percent} of its current score for each further percent
by which they exceed the best submitted solution; teams submitting a solution with $B(I)$ colors
receive a score of $1$. Teams that did not submit any valid solution for an instance
receive a default score of $0$, corresponding to a solution with infinitely many colors.

The total score of each team $T$ was then calculated by summing the score for that team over all instances $I$.
The winner of the contest was the team with the highest score.
In case of ties, the tiebreaker was set to be the time a specific score was obtained. 
This turned out to be unnecessary.

\subsection{Categories}
The contest was run in an \emph{Open Class}, in which participants could use any
computing device, any amount of computing time (within the duration of the
contest) and any team composition. In the \emph{Junior Class}, a
 team was required to consist exclusively of participants who were eligible
according to the rules of CG:YRF (the \emph{Young Researchers Forum} of CG Week), 
defined as not having defended a formal doctorate before 2020.

\subsection{Server and Timeline}
The contest itself was run through a dedicated server at TU Braunschweig,
hosted at \url{https://cgshop.ibr.cs.tu-bs.de/competition/cg-shop-2022/}.
It opened at 00:00 (UTC) on September 19, 2021, and closed
at 24:00 (midnight, AoE), on January 19, 2022. 

\section{Outcomes}
A total of 40 teams signed up for the competition, and 32 teams submitted
at least one valid solution.
In the end, the leaderboard for the top 10 teams looked as shown in Table~\ref{tab:top10}.
The overall winner of the contest are Team Shadoks, with a perfect score of 225, meaning that
no other team beat their solution on any instance.
\begin{table}[h!]
	\centering
	\begin{tabular}{|r|l|r|r|}
		\hline
		\textbf{Rank} & \textbf{Team} & \textbf{Score} & \textbf{Junior} \\
		\hline
		1 &	Shadoks & 225.000 & \\
		2 & gitastrophe & 217.486 & \checkmark\\
		3 & LASAOFOOFUBESTINNRRALLDECA & 211.803 & \\
		4 & TU Wien & 195.967 & \checkmark\\
		5 & OMEGA-UHA & 144.481 & \\
		6 & Luiwammus & 131.275 & \checkmark\\
		7 & Expressway65 & 127.765 & \\
		8 & The Turtles & 123.600 & \checkmark\\
		9 & Wouter & 122.317 & \checkmark\\
		10 & ALGA@TUe & 119.213 &\\
		\hline
	\end{tabular}
	\caption{The top 10 of the final score, rounded to three decimal places.
	         Teams that satisfy the criteria for being considered a junior team
			 have a checkmark in the ``Junior'' column.}
	\label{tab:top10}
\end{table}

\begin{figure}
  \centering
  \includegraphics[width=.98\textwidth]{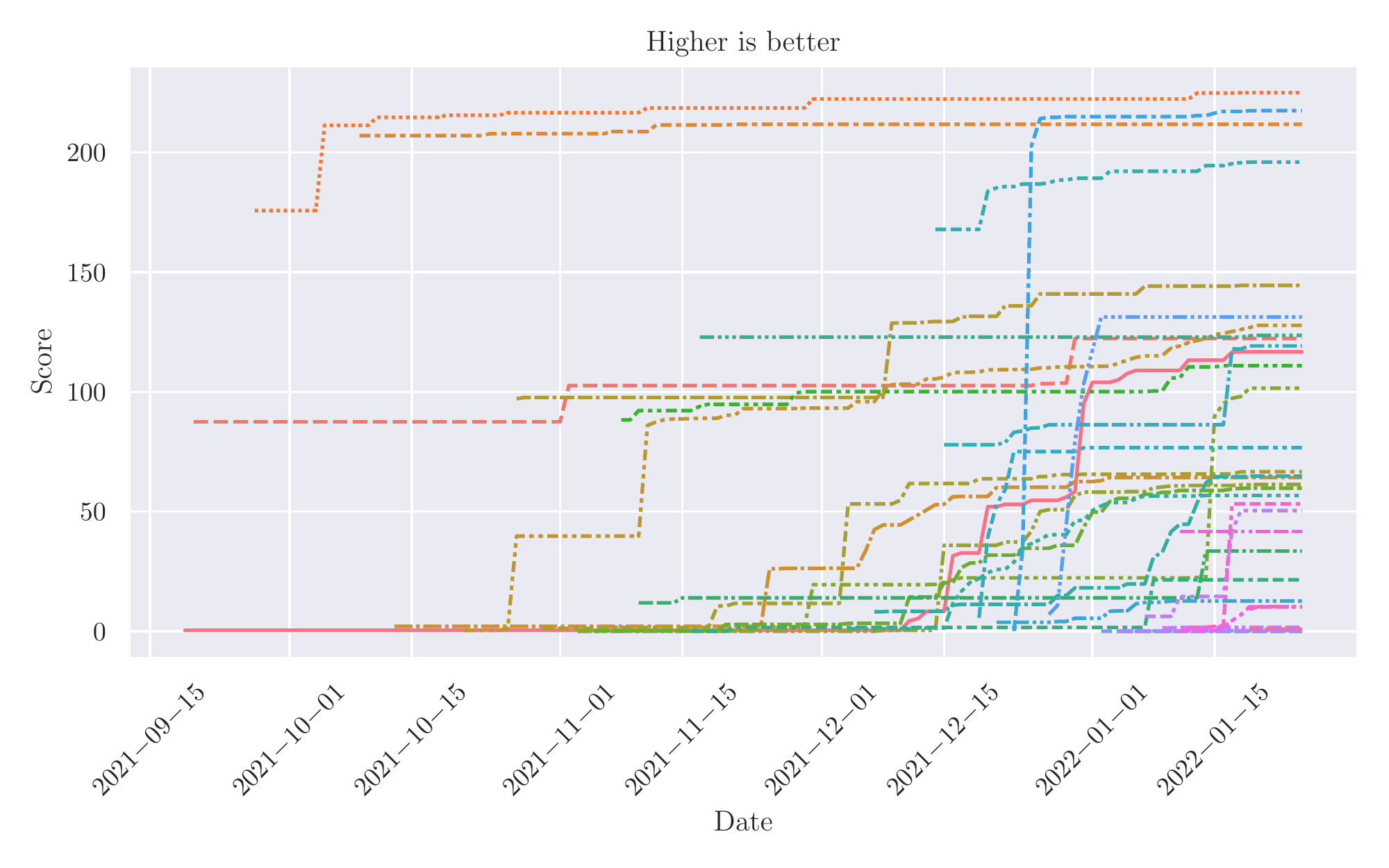}
  \caption{Score progress of the teams over time.
  It is remarkable that the initial solutions submitted by the team Shadoks in September were only surpassed by three other teams.
  By multiple smaller improvements, team Shadoks maintained their lead after their initial submission until the end.
  Team LASAOFOOFUBESTINNRRALLDECA also joined early and strong while team gitastrophe only joined in late December, quickly raising to rank 2.}
\label{fig:score-progress}
\end{figure}

\begin{figure}
  \centering
  \includegraphics[width=.98\textwidth]{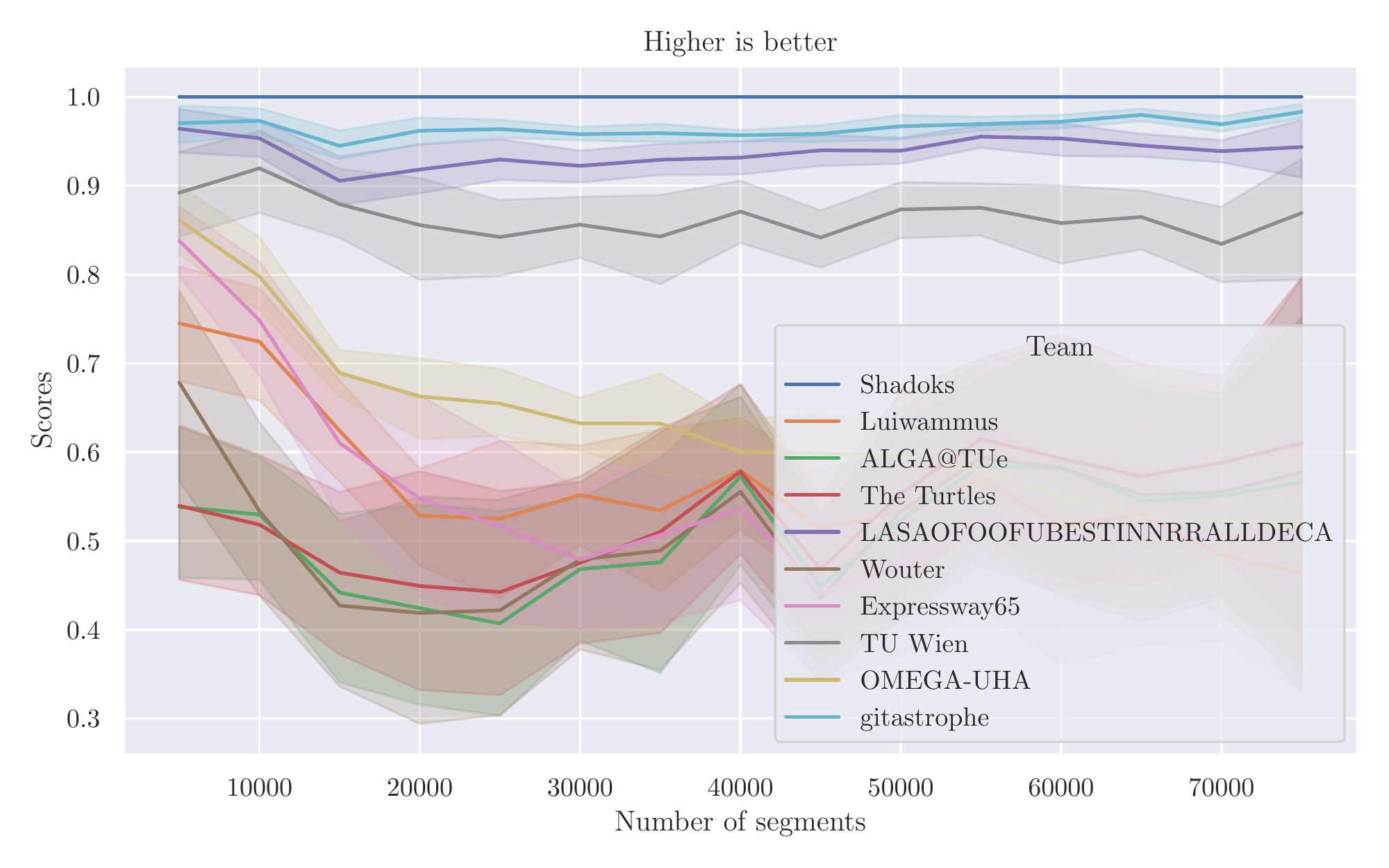}
  \caption{The (mean) scores over the instance sizes of the ten best teams. We can see
           that team Shadoks submitted the best solution on all instances, leading to the perfect score of 225.
           The performance of team TU Wien seems to slightly decrease with size, while the top 3 teams get closer.
           Most other teams reach slightly higher scores for the large instances than for the medium-sized instances.}
\label{fig:scores_over_size}
\end{figure}

The progress over time of each team's score can be seen in Figure~\ref{fig:score-progress};
the best solutions for all instances (displayed by score) can be seen in Figure~\ref{fig:scores_over_size}.
The top four teams were invited for contributions in the 2022 SoCG proceedings, as follows.
\begin{enumerate}
\item Team Shadoks: Loïc Crombez, Guilherme D. da Fonseca, Yan Gerard and Aldo Gonzalez-Lorenzo~\cite{shadoks}.
\item Team gitastrophe: Jack Spalding-Jamieson, Brandon Zhang and Da Wei Zheng~\cite{gitastrophe}.
\item Team LASAOFOOFUBESTINNRRALLDECA: Florian Fontan, Pascal Lafourcade, Luc Libralesso and Benjamin Momège~\cite{lasawhatever}.
\item Team TU Wien: André Schidler~\cite{tuwien}.
\end{enumerate}

All these teams engineered their solutions based on a spectrum of heuristics for generating initial solutions.
Some teams relied on well-known graph coloring heuristics for the initial solutions, some augmented these heuristics,
making use of the geometric nature of the problem.
These initial solutions were then improved by various methods, including local search;
several teams based their improvement search on an adaptation of the conflict optimization method
used by Team Shadoks in the previous installment of the contest~\cite{crombez2021shadoks}.
Details of their methods and the engineering decisions they made are given in their respective papers.

For comparison, we ran the heuristics DSATUR and RLF, as well as several hundred iterations of the iterated greedy improvement strategy
on all instances; we also ran a previously existing implementation~\cite{lewisbook} of a simple,
tabu-search-based hybrid evolutionary algorithm on all instances with a very modest time limit of $2$ minutes.
Table~\ref{tab:portfolio} shows how these algorithms would have performed in the contest.

\begin{table}[h!]
	\centering
	\begin{tabular}{|r|l|l|}
		\hline
		\textbf{Heuristic} & \textbf{Score} & \textbf{Rank}\\
		\hline
		Hybrid EA & 61.017 & 19\\
		Iterated Greedy & 59.743 & 19\\
		DSATUR \& RLF & 53.924 & 21\\
		DSATUR & 52.809 & 22\\
		RLF & 33.698 & 24\\
		\hline
	\end{tabular}
	\caption{The score that simple heuristics and metaheuristics would have achieved,
	         and the rank they would have gotten, had they participated in the contest.}
	\label{tab:portfolio}
\end{table}

This shows the massive advantage of the approaches of the top teams over simple, out-of-the-box methods;
it also shows that the majority of contest participants actually came up with algorithms that 
were able to beat such methods.

\section{Conclusions}

The 2022 CG:SHOP Challenge motivated a considerable number of teams to engage in extensive optimization studies.
The outcomes promise further insight into the underlying, important optimization problem.
Moreover, the success of the teams that based their approach on the winning team's approach from last year,
which considered a completely different problem, shows that the contest may drive research that may be applicable to
various problems from different problem domains.
Moreover, the considerable participation of junior teams indicates that the Challenge itself
motivates a great number of students and young researchers to work on practical algorithmic problems.

\bibliography{bibliography}
\end{document}